\documentstyle[aps,epsf,floats]{revtex}
\begin{document}
\newcommand\1{$\spadesuit$}
\newcommand\2{$\clubsuit$}
\draft
\twocolumn[\hsize\textwidth\columnwidth\hsize\csname 
@twocolumnfalse\endcsname
\title{The Robustness of Inflation to Changes in Super-Planck-Scale Physics}

\author{Robert~H.~Brandenberger}
\address{Department of Physics, Brown University, Providence, RI 02912, USA.\\
e-mail: rhb@het.brown.edu
}
\author{J\'er\^ome Martin}
\address{Institut d'Astrophysique de Paris, \\ 
98 boulevard Arago, 75014 Paris, France. \\
e-mail: jmartin@ccr.jussieu.fr
}

\date{April 3, 2001}
\maketitle

\begin{abstract}
We calculate the spectrum of density fluctuations in models of inflation based 
on a weakly self-coupled scalar matter field minimally coupled to gravity, and 
specifically investigate the dependence of the predictions on modifications of 
the physics on length scales smaller than the Planck length. These 
modifications are encoded in terms of modified dispersion relations. Whereas 
for some classes of dispersion relations the predictions are unchanged 
compared to the usual ones which are based on a linear dispersion 
relation, for other classes important differences are obtained, involving 
tilted spectra, spectra with exponential factors and with 
oscillations. This is the case when the dispersion relation becomes 
complex. We conclude that the predictions of inflationary cosmology 
in these models are not robust against changes in the 
super-Planck-scale physics.
\end{abstract}

\pacs{PACS numbers: 98.80.Cq, 98.70.Vc}
\vspace*{1cm}
]

\section{Introduction}

Most current models of inflation \cite{Guth} are based on weakly self-coupled 
scalar matter fields minimally coupled to gravity. In most of these models,
the period of inflation lasts for a number of e-foldings much larger than the 
number needed to solve the problems of standard cosmology \cite{Linde}. In 
these cases, the physical length of perturbations of cosmological interest 
today (those which today correspond to the observed CMB anisotropies and 
to the large-scale structure) was much smaller than the Planck length at 
the beginning of inflation. Hence, the approximations which go into the 
calculation of the spectrum of cosmological perturbations \cite{flucts} 
break down. It is then of interest to investigate whether the predictions 
are sensitive to the unknown super-Planck-scale physics, or whether the 
resulting spectrum of perturbations is determined only by infrared physics.

An analogous problem arises for black hole evaporation. The original 
computations of the thermal spectrum from black holes \cite{Hawking} 
appear to involve mode matching at super-Planck scales. However, in the 
case of black holes it can be shown \cite{Unruh,Brout,Hambli,Corley} that 
the predictions are in fact insensitive to modifications of the physics 
at the ultraviolet end. 

Our goal is to explore whether and in which cases the spectrum of 
fluctuations resulting from inflationary cosmology depends on the 
unknown ultraviolet physics. We will adapt the method of 
\cite{Unruh,Corley} and consider theories obtained by replacing the 
linear dispersion relation for the linearized fluctuation equations 
by classes of nonlinear dispersion relations. We find that for the 
class of dispersion relations introduced by Unruh \cite{Unruh} one 
recovers a scale-invariant spectrum of fluctuations in the case of 
exponential inflation. In contrast, for the class of dispersion 
relations modelled after the one introduced in \cite{Corley}, the 
resulting spectrum may be tilted and may include exponential and 
oscillatory factors if the dispersion relation becomes complex. Such 
spectra are inconsistent with 
observations. We thus conclude that the predictions for observables 
in weakly coupled scalar field models of inflation depend 
sensitively on hidden assumptions about super-Planck-scale physics. 
  
\section{Framework}

We will consider the evolution of linear cosmological fluctuations in 
a spatially flat homogeneous and isotropic Universe. As is well known 
(see e.g. \cite{MFB92} for a comprehensive review), the evolution 
equation of scalar and tensor fluctuations in conformal time $\eta$ 
reduces to harmonic oscillator equations with time dependent masses. In 
the following, we will therefore simply consider the 
evolution of a scalar field $\Phi(\eta, \bf{x})$ living 
on space-time. Introducing $\mu(n,\eta )$ via the Fourier transform 
$\Phi (\eta ,{\bf x}) \equiv [1/(2\pi )^{3/2}]
\int {\rm d}{\bf n}(\mu /a)e^{i{\bf n}\cdot {\bf x}}$ 
[where $a(\eta)$ is the scale factor], the evolution equation of 
the mode with a comoving wavenumber $n$ becomes
\begin{equation} \label{eom}
\mu ''+\biggl[n^2-\frac{a''}{a}\biggr]\mu \, = \, 0 \, .
\end{equation} 
The corresponding power spectrum $P_{\Phi}(n)$ is given by  
\begin{equation} \label{spec}
n^3P_{\rm \Phi } \, = \, n^3\biggl \vert \frac{\mu }{a}\biggr \vert ^2 \, .
\end{equation}

In order to study the dependence of the predictions for $P_{\Phi}$ on 
super-Planck-scale physics, we will modify the linear dispersion relation 
$\omega_{\rm p}^2 = k^2 = (n/a)^2$ (where $\omega_{\rm p}$ is the physical frequency) 
for wavenumbers greater than a critical wavenumber $k_{\rm C}$ by replacing 
the $n^2$ term in (\ref{eom}) with  
\begin{equation}
\label{neff}
n_{\rm eff}^2 \, = \, a^2(\eta )F^2(k) \, = \, a^2(\eta )F^2[n/a(\eta)] \, ,
\end{equation}
where $F(k)$ differs significantly from $k$ only for $k > k_{\rm C}$. We see 
that, in terms of comoving wavenumbers, we obtain a time 
dependent dispersion relation. 

The two classes of dispersion relations we
specifically analyze are the one proposed by Unruh \cite{Unruh} and a 
generalization of the one studied by Corley and Jacobson \cite{Corley}. The 
first class is given by 
\begin{equation}
\label{Udisp}
F(k) \, \equiv \, k_{\rm C}\tanh ^{1/p}\biggl
[\biggl(\frac{k}{k_{\rm C}}\biggr)^p\biggl],
\end{equation}
where $p$ is an arbitrary coefficient.
For large values of the 
wave number, this becomes a constant $k_{\rm C}$ 
whereas for small values this is a linear law as expected.
The second class of dispersion relations is given by
\begin{equation}
\label{Jdisp}
F^2(k) \,= \, k^2 + k^2 b_m\biggl(\frac{k}{k_{\rm C}}\biggr)^{2m},
\end{equation}
where $m$ is an integer and the coefficients $b_m$ are at this stage 
arbitrary. Note that for negative $b_m$ the dispersion relation becomes 
complex for $k \gg k_{\rm C}$. The dispersion relations are shown in Fig. 1.

\begin{figure}
\begin{center}
\leavevmode
\hspace*{-2.1cm}
\epsfxsize=10.5cm
\epsffile{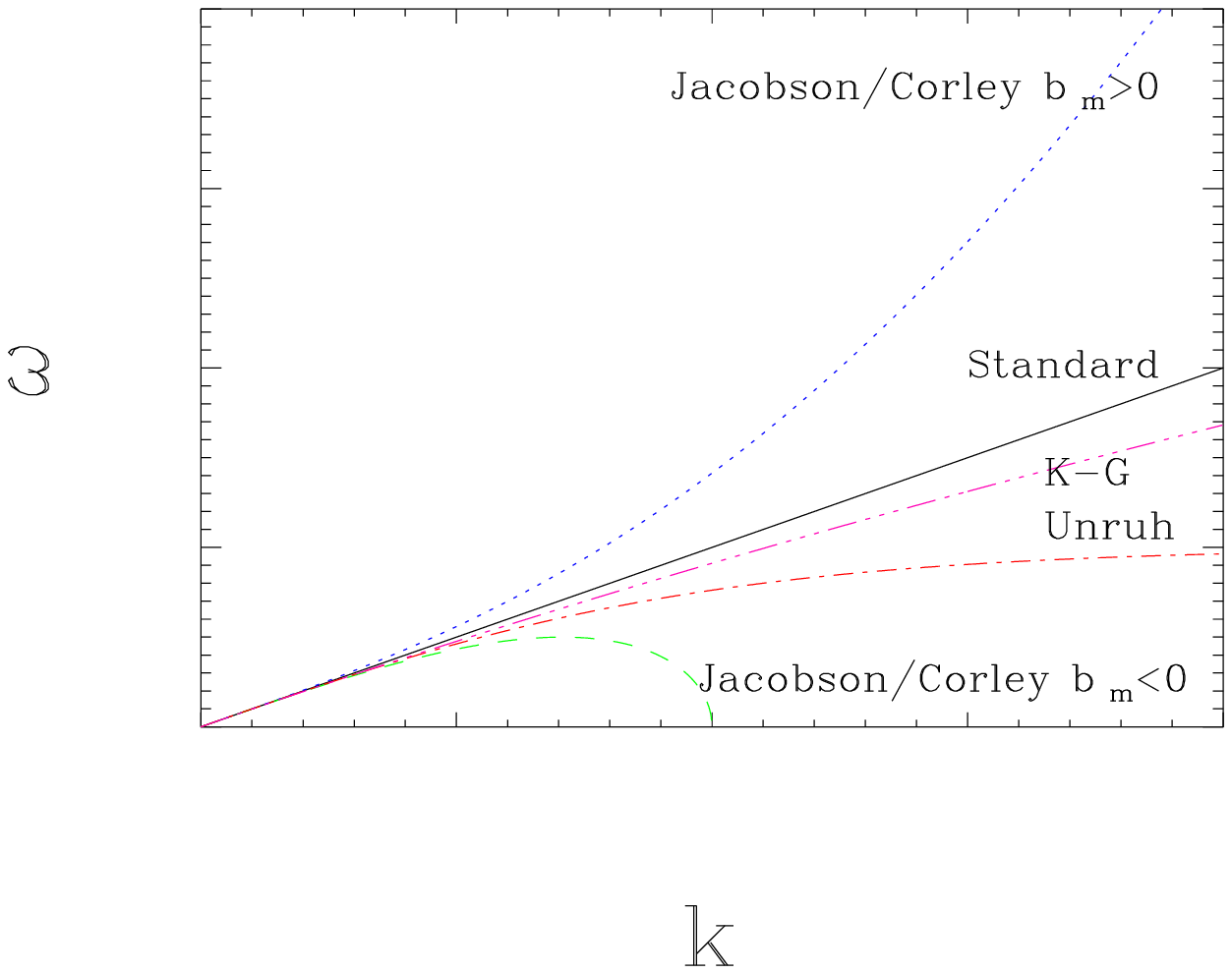}
\end{center}
\caption{Sketch of the different dispersion relations.}
\label{disp}
\end{figure}

In order to compute the power spectrum $P_{\Phi}(n)$ of (\ref{spec}) we need 
to know both the initial conditions for the mode $\mu(n,\eta )$ and the subsequent 
evolution. We first discuss the initial conditions. We want the state at the 
initial time $\eta_{\rm i}$ to correspond as closely as possible to our usual 
physical intuition of a vacuum state. We here study two prescriptions for 
this. The first is to canonically quantize the field $\mu (n,\eta )$ and to demand 
that the initial state minimizes the energy \cite{Brown78}. The second is 
to set the state up in the local Minkowski vacuum \cite{RB84}. In the case 
of a linear dispersion relation, both prescriptions give the same 
result. However, for a nonlinear dispersion relation, we obtain different 
results (which emphasizes the general point that the predictions of 
inflationary cosmology depend on the initial state chosen \cite{MRS99}). 

Demanding that the initial state minimize the energy yields \cite{MB00} 
\begin{equation} \label{inimu}
\mu (\eta =\eta _{\rm i}) \, =\, \frac{1}{\sqrt{2\omega (\eta _{\rm i})}}, 
\quad  
\mu '(\eta =\eta _{\rm i}) \, = \, \pm 
i\sqrt{\frac{\omega (\eta _{\rm i})}{2}}\, , 
\end{equation}
where $\omega$ is the comoving frequency, 
whereas prescribing the local Minkowski vacuum state gives
\begin{equation} \label{Minvac}
\mu (\eta _{\rm i}) \, = \, \frac{1}{\sqrt{2n}}, \quad \mu '(\eta _{\rm i}) \, =
\, \pm i\sqrt{\frac{n}{2}} \, .
\end{equation}
As is apparent from (\ref{eom}), in the case of a linear dispersion 
relation $\omega = n$ and the two prescriptions for the initial state 
coincide. It should be clear that the choice (\ref{inimu}) is
certainly the most physical one. The second choice, as mentioned
above, illustrates the fact that the final result does depend on the 
initial conditions.

\section{Calculation of Spectra}

We now compute the power spectrum $P_{\Phi}(n)$ for values of $n$ with 
wavenumbers larger than $k_{\rm C}$ at the initial time $\eta_{\rm i}$. On such 
scales, the time interval can be divided into three regions. The 
first is $\eta_{\rm i} < \eta < \eta_1(n)$, during which the physical wavenumber 
exceeds $k_{\rm C}$. During this time interval, the mode evolution is 
non-standard. The second interval lasts from the time $\eta_1(n)$ 
to the time $\eta_2(n)$ when the wavelength equals the 
Hubble radius $l_{\rm H}$. During this time, the solutions of the mode 
equation (\ref{eom}) are oscillatory since the $n^2$ term in the 
parentheses in (\ref{eom}) dominates over the $a'' / a$ term. In the 
third period [$\eta > \eta_2(n)$], the modes are effectively 
frozen: the non-decaying solution of the mode 
equation is $\mu(\eta) \sim a(\eta)$. 


The values of $\eta_1$ and $\eta_2$ depend on the background 
evolution. For a power-law inflation model 
with $a(\eta) = l_0\vert \eta \vert ^{1+\beta }$, where $\beta $ is 
a number with $\beta \le -2$ \footnote{The value $\beta = -2$ 
corresponds to exponential inflation.} and $l_0$ has the dimension 
of a length, the values of $\eta_1$ and $\eta_2$ are given by
\begin{eqnarray} 
\label{time1}
\vert \eta _1(n)\vert \, &=& \, 
\biggl(\frac{n}{2\pi }\frac{l_{\rm C}}{l_0}\biggr)^{1/(1+\beta )} 
\vert b_m\vert ^{1/[2m(1+\beta )]}\, , 
\\
\vert \eta _2(n)\vert &=& \frac{2\pi}{n} \vert 1+\beta \vert \, ,
\end{eqnarray}
where $l_{\rm C}$ is the wavelength corresponding to $k_{\rm C}$. By 
combining the above formula for $a(\eta)$ with (\ref{time1}), it follows that
\begin{equation} \label{scale2}
a[\eta_2(n)]^{-2} \, \sim \, n^{2 + 2 \beta} \, .
\end{equation}
Assuming that the non-decaying modes mix with coefficients of order 
unity at the times $\eta_1$ and $\eta_2$, then based on the above 
observations about the time dependence of $\mu(n,\eta)$ in the various 
time intervals, we obtain the following `master formula' for the 
power spectrum at late times:
\begin{equation} \label{master}
n^3 P_{\Phi}(n) \, \sim \, {{n^3} \over {2 \omega(\eta_{\rm i})}} 
\biggl\vert {{\mu [n,\eta_1(n)]} \over {\mu(n,\eta_{\rm i})}}\biggr \vert^2 
a[\eta_2(n)]^{-2} \, .
\end{equation} 
This result is true if the initial state minimizes the energy density. If 
the initial state is taken to be the local Minkowski vacuum, then 
$\omega(\eta_{\rm i})$ on the r.h.s. of (\ref{master}) needs to be replaced by $n$.

In the case of the {\bf linear dispersion relation}, $\mu(n,\eta)$ also 
oscillates during the first time interval $\eta_{\rm i} < \eta < \eta_1(n)$. Since 
in this case $\omega(\eta_{\rm i}) \sim n$, we immediately obtain
\begin{equation} \label{power1}
n^3 P_{\Phi}(n) \, \sim \, n^{4 + 2 \beta} \, ,
\end{equation}
the `standard' prediction of inflationary cosmology.

In the case of {\bf Unruh's dispersion relation}, the mode equation can 
be solved exactly during the first time interval in the case of 
exponential inflation ($\beta = -2$):
\begin{equation} \label{solU}
\mu (n,\eta ) \, = \, A_1\vert \eta \vert ^{x_1}+A_2\vert \eta \vert ^{x_2} \, ,
\end{equation}
where $A_1$ and $A_2$ are two constant determined by the initial 
conditions and where the exponents $x_1$ and $x_2$ are given by
\begin{equation} \label{expU}
x_{1,2} \, \equiv \, \frac{1}{2}\pm 
\frac{1}{2}\sqrt{9-16\pi ^2\frac{l_0^2}{l_{\rm C}^2}}
\, .
\end{equation}
Note that both modes are 
decaying ($\vert \mu \vert \sim \eta^{1/2}$ and 
$\vert \eta \vert \rightarrow 0$). Since $\eta_1$ depends on $n$ 
as given in (\ref{time1}), we have
\begin{equation}
{{\mu[n,\eta_1(n)]} \over {\mu(n,\eta_{\rm i})}} \, \sim \, n^{-1/2} \, .
\end{equation}
Since for the minimum energy density initial state $\mu(n,\eta_{\rm i})$ is 
independent of $n$ in the wavelength interval under consideration, we obtain
\begin{equation}
n^3 P_{\Phi}(n) \, \sim \, n^0 \, ,
\end{equation}
i.e. the same scale-invariant spectrum as in the case of the linear 
dispersion relation.

\begin{table*}
\caption{Spectra for different dispersion relations and different initial 
conditions. The function $\bar{B}(n)$ is a complicated oscillatory 
functions which can be found in Ref. [13].}
\label{summary}
\begin{tabular}{cccccc}
 
Dispersion Relation & Initial Conditions  & Spectrum  $n^3P_{\rm \Phi}$ \\ \hline 
\\
Unchanged \quad ($\omega =k$) & Minimizing energy=Minkowski  & $n^{2\beta +4}$   \\

Unruh \quad ($\beta =-2$) & Minimizing energy & $n^0$   \\

Unruh \quad ($\beta =-2$) & Minkowski  & $n^{-1}
\cos ^2\biggl(\frac{2\pi }{\epsilon }+\frac{2\pi }{\epsilon }\ln \biggl 
\vert \frac{2\pi }{n\eta _{\rm i}}\biggr \vert \biggr)$       \\

Jacobson/Corley \quad ($b_{\rm m}<0$)
& Minimizing energy & 
$n^{2\beta +4}e^{An^{m+1}}\cos ^2\biggl[2\pi |1+\beta | 
-\biggl(\frac{\epsilon }{2\pi }\biggr)^{\frac{1}{1+\beta }}n^{\frac{2+\beta }{1+\beta }}
-\frac{\pi }{4}\biggr] $ \\

Jacobson/Corley \quad ($b_{\rm m}<0$) & Minkowski & 
$n^{2\beta +4+m}e^{An^{m+1}}\cos ^2\biggl[2\pi |1+\beta | 
-\biggl(\frac{\epsilon }{2\pi }\biggr)^{\frac{1}{1+\beta }}n^{\frac{2+\beta }{1+\beta }}
-\frac{\pi }{4}\biggr] $ \\

Jacobson/Corley \quad ($b_{\rm m}>0$) & Minimizing energy  & 
$n^{2\beta +4}$ \\

Jacobson/Corley \quad ($b_{\rm m}>0$) & Minkowski & 
$n^{2\beta +4+m}\vert \bar{B}(n)\vert ^2$ \\

\end{tabular}

\end{table*}



In the case of the {\bf Corley/Jacobson dispersion relation} the result 
is quite different. Let us consider the case $b_m < 0$ (complex 
dispersion relation). Then, in the wavelength regime 
$\lambda(\eta_{\rm i}) \ll l_{\rm C}$ of interest, the mode equation in the first 
time interval can be solved exactly in terms of modified Bessel functions
\begin{equation} \label{modeCJ}
\mu(n,\eta) \, \sim \, \vert \eta \vert^{1/2} I_{1/(2b)}(z) \, \sim \,
\vert \eta \vert^{1/2} [z(\eta)]^{-1/2} e^{z(\eta)} \, ,
\end{equation}
where the argument of the Bessel function is 
$z(\eta) \equiv \gamma \vert \eta \vert^b$ with 
\begin{equation}
b \, \equiv  \, 1 - m(1 + \beta), \quad 
\gamma \,\equiv  \, {{\sqrt{\vert b_m \vert}} \over {(2 \pi)^m b}} 
\biggl({{l_{\rm C}} \over {l_0}}\biggr)^m n^{m + 1} \, .\nonumber
\end{equation}
Note, in particular, the exponential factor in (\ref{modeCJ}) which 
depends on $n$. This factor does not cancel with any other $n$-dependent 
term in (\ref{master}) and is thus the root of the exponential 
dependence of the final power spectrum on $n$. Combining 
(\ref{master}), (\ref{inimu}), (\ref{scale2}) and (\ref{modeCJ}) we obtain
\begin{eqnarray}
n^3 P_{\Phi}(n) \, &\sim& \, n^3 n^{-1 - m} n^m e^{2 z[\eta_1(n)]} 
n^{2 + 2 \beta} \nonumber \\
&\sim& n^{4 + 2 \beta} e^{2 z[\eta_1(n)]} \, ,
\end{eqnarray}
where the second factor on the r.h.s. of the first line comes from 
$\omega(\eta_{\rm i})$, the third and fourth factors stem from the ratio 
of $\mu$, and the final factor from the $a(\eta_2)$ term in 
(\ref{master}). The careful matching between growing and decaying
modes also reveals \cite{MB00} the presence of an oscillating factor 
$\cos ^2(n\eta_2 - n\eta_1 - \pi /4)$
in the final power spectrum. However, it should also be noticed that 
the initial conditions are fixed in a region where the mode function 
does not oscillate. 
\par
In the case of the Corley-Jacobson dispersion relation with $b_m > 0$, the 
modified Bessel functions in the mode equation during the first time 
interval must be replaced by regular Bessel functions. Hence, the 
exponential factors in the power spectrum disappear and the final 
result is unchanged, i.e. we recover a scale invariant spectrum. 

\section{Discussion and Conclusions}

We have studied the robustness of the predictions for the spectrum of 
cosmological perturbations of weakly coupled inflationary models. The 
method used was to replace the usual linear dispersion relation by 
special classes of nonlinear ones, where the nonlinearity is confined to 
physical wavelengths $\lambda$ smaller than some critical length $l_{\rm C}$. We 
found that for the class of dispersion relations first introduced by 
Unruh \cite{Unruh}, the predictions are unchanged. This is connected 
with the fact that the initial vacuum state evolves adiabatically up to 
the time $\eta_1$ when $\lambda = l_{\rm C}$ \footnote{We thank Bill Unruh for 
pointing this connection out to us.}. However, in the case of the dispersion 
relation modelled after the one used by Corley and Jacobson 
\cite{Corley} in the situation where it becomes complex, the resulting 
spectrum can have oscillations, non-standard 
tilts and exponential factors which render the resulting theory in 
conflict with observations. The specific predictions depend on the 
sign of $b_m$, on the value of $m$, and on the initial state chosen. The 
results are summarized in Table 1.

We thus conclude that the predictions in weakly coupled scalar field-driven 
inflationary models are not robust to changes in the unknown fundamental 
physics on sub-Planck lengths. This opens up another potentially very 
interesting link between fundamental physics and observations. Note, 
however, that in strongly coupled scalar field models of inflation 
such as the model discussed in \cite{BZ}, the spectrum of fluctuations 
is robust to changes in the underlying sub-Planck-length physics.

\vspace{0.5cm} 
\centerline{\bf Acknowledgements}
\vspace{0.2cm}

We are grateful to Lev Kofman, Dominik Schwarz, Carsten Van de Bruck 
and in particular Bill Unruh for 
stimulating discussions and useful comments. We acknowledge support 
from the BROWN-CNRS University Accord which made possible the visit of J.~M. to 
Brown during which most of the work on this project was done, and we are 
grateful to Herb Fried for his efforts to secure this Accord. One of us (R.~B.) 
wishes to thank Bill Unruh for hospitality at the University of British Columbia 
during the time when this work was completed. J.~M. thanks the High Energy 
Group of Brown University for warm hospitality. The research was supported in 
part by the U.S. Department of Energy under Contract DE-FG02-91ER40688, TASK A.


\begin{references}

\bibitem{Guth} A.~Guth, {\it Phys. Rev.} {\bf D23}, 347 (1981).

\bibitem{Linde} A.~Linde, D.~Linde and A.~Mezhlumian, {\it Phys. Rev.} 
{\bf D49}, 1783 (1994); \\
A.~Linde, `Lectures on Inflationary Cosmology', Stanford preprint SU-ITP-94-36,
{\tt hep-th/9410082} (1994).

\bibitem{flucts} G.~Chibisov and V.~Mukhanov, `Galaxy Formation and
Phonons,' Lebedev Physical Institute Preprint No. 162 (1980);\\
V.~Mukhanov and G.~Chibisov, {\it JETP Lett.} {\bf 33}, 532 (1981);\\
V.~Mukhanov and G.~Chibisov, {\it Sov. Phys. JETP} {\bf 56}, 258 (1982);\\
G.~Chibisov and V.~Mukhanov, {\it Mon. Not. R. Astron. Soc.} {\bf
200}, 535 (1982);\\
V.~Lukash, {\it Pis'ma Zh. Eksp. Teor. Fiz.} {\bf 31}, 631 (1980);\\ A.~Starobinsky, {\it Phys. Lett.} {\bf 117B}, 175 (1982);\\
S.~Hawking, {\it Phys. Lett.} {\bf 115B}, 295 (1982);\\
A.~Guth and S.~Y.~Pi, {\it Phys. Rev. Lett.} {\bf 49}, 1110 (1982);\\
J.~Bardeen, P.~Steinhardt, and M.~Turner, {\it Phys. Rev.} 
{\bf D28}, 679 (1983).

\bibitem{Hawking} S.~Hawking, {\it Comm. Math. Phys.} {\bf 43}, 199 (1975).
 
\bibitem{Unruh} W.~Unruh, {\it Phys. Rev.} {\bf D51}, 2827 (1995).

\bibitem{Brout} R.~Brout, S.~Massar, R.~Parentani and P.~Spindel, 
{\it Phys. Rev.} {\bf D52}, 4559 (1995).

\bibitem{Hambli} N.~Hambli and C.~Burgess, {\it Phys. Rev.} {\bf D53}, 5717 (1996)

\bibitem{Corley} S.~Corley and T.~Jacobson, {\it Phys. Rev.} {\bf D54}, 1568 (1996);\\
S.~Corley, {\it Phys. Rev.} {\bf D57}, 6280 (1998).

\bibitem{MFB92} V.~Mukhanov, H.~Feldman, and R.~Brandenberger, {\it Phys. 
Rep.} {\bf 215}, 203 (1992).
 
\bibitem{Brown78} M.~Brown and C.~Dutton, {\it Phys. Rev.} {\bf D18}, 4422 (1978).

\bibitem{RB84} R.~Brandenberger, {\it Nucl. Phys.} {\bf B245}, 328 (1984).

\bibitem{MRS99} J.~Martin, A.~Riazuelo and M.~Sakellariadou, {\it Phys. Rev.} 
{\bf D61}, 083518 (2000). {\tt astro-ph/9904167}.

\bibitem{MB00} J. Martin and R. Brandenberger, `The Trans-Planckian Problem of Inflationary Cosmology', Brown preprint BROWN-HET-1214.

\bibitem{BZ}  R.~Brandenberger and A.~Zhitnitsky, {\it Phys. Rev.} 
{\bf D55}, 4640 (1997).



 
\end{references}
\end{document}